\documentclass[11pt,a4paper]{article}
\pdfoutput=1
\usepackage{jheppub}
\usepackage{amsmath}
\usepackage{caption}
\usepackage{subcaption}

\newcommand{\fig}[1]{Fig.\ \ref{#1}}

\newcommand{\eqq}[1]{Eq.~(\ref{#1})}

\newcommand{\be}{\begin{equation}}
\newcommand{\ee}{\end{equation}}

\newcommand{\bea}{\begin{eqnarray}}
\newcommand{\eea}{\end{eqnarray}}

\def\phiM{\phi_M}

\begin{document}
\begin{titlepage}
\thispagestyle{empty}

\vspace{2cm}
\begin{center}
\font\titlerm=cmr10 scaled\magstep4
\font\titlei=cmmi10
scaled\magstep4 \font\titleis=cmmi7 scaled\magstep4 {\Large{\textbf{Holographic drag force in non-conformal plasma}
\\}}
\setcounter{footnote}{0}
\vspace{1.5cm} \noindent{{
Tolga Domurcukgul$^a$\footnote{e-mail:tolga.domurcukgul@boun.edu.tr}, and
Razieh Morad$^{a,b}$\footnote{e-mail:r.morad@ipm.ir} 
}}\\
\vspace{0.8cm}

{\it $^a$ Department of Physics, Bo\u{g}azi\c{c}i University, Istanbul, Turkey \\}
{\it $^b$ School of Particles and Accelerators, Institute for Research in Fundamental Sciences (IPM), Tehran, Iran}

\vspace*{.4cm}
\end{center}
\vskip 2em
\setcounter{footnote}{0}
\begin{abstract}

In this study, the gauge/string duality is used to investigate the dynamics of a moving heavy quark in a strongly coupled, non-conformal plasma. The drag force in this non-conformal model is smaller than that of $\mathcal{N} = 4$ supersymmetric Yang-Mills (SYM) plasma and decreases as the level of non-conformality is increased.  At intermediate temperatures, the world-sheet temperature, which is derived numerically by calculating the world-sheet horizon, deviates from its conformal value, but at high temperatures, it tends to its conformal value.
\end{abstract}
\tableofcontents
\end{titlepage}
\section{Introduction}

Heavy-ion collisions at the Relativistic Heavy Ion Collider (RHIC) and the Large Hadron Collider (LHC) have provided strong evidence indicating the formation of the strongly coupled quark-gluon plasma (QGP), a deconfined state of hadronic matter whose dynamics after the collision is dominated by non-perturbative effects \cite{Adams:2005dq,Adcox:2004mh,Arsene:2004fa,Back:2004je,Policastro:2001yc,Kovtun:2004de,Baier:1996kr,Eskola:2004cr}. While the perturbative QCD works only in the weak coupling regime, the string/gauge duality can be used as the powerful tool fo study the strongly coupled plasma \cite{Maldacena:1997re,Witten:1998qj,Gubser:1998bc,Aharony:1999ti,CasalderreySolana:2011us} .

However, there is no known gravity dual to QCD, the duality between the $\mathcal{N} = 4\, SU(N_c)$ super-Yang-Mills theory and type IIB string theory on $AdS_5 \times S^5$ is the most studied example in the context of AdS/CFT correspondence \cite{Shuryak:2008eq,Shuryak:2004cy,Baier:1996kr} which provided promising results. For example, the ratio of shear viscosity over entropy density obtained from this duality, $\eta/s = 1/4\pi$ \cite{Policastro:2001yc,Kovtun:2003wp,Buchel:2003tz} is consistent with the experimental data \cite{Teaney:2003kp}. 

Lattice data suggests that the QGP formed at high-energy heavy-ion collisions is not a fully conformal fluid, and bulk viscosity, which is a purely non-conformal effect, is needed for the precise extraction of the shear viscosity of the QGP \cite{Ryu:2015vwa}. Hydrodynamics including non-conformal effects successfully describes the smaller system such as p-Pb \cite{Bozek:2011if} and p-p \cite{Schenke:2014zha,Habich:2015rtj} collisions \cite{Jeon:2015dfa}. The original duality can be extended to the theories more similar to QCD or QGP using the well-known top-down \cite{Polchinski:2000uf,Karch:2002sh,Sakai:2004cn} or a bottom-up \cite{Gursoy:2007cb,Galow:2009kw} approaches. 

One example of the latter approach is a five-dimensional gravity model coupled to a scalar field with a non-trivial potential \cite{Attems:2016ugt}. In this model, the conformal invariance breaks even at zero temperature by coupling a scalar field at pure gravity in AdS, which duals to a CFT deformed by a source $\Lambda$ for a dimension-three operator. This source breaks the scale invariance of the resulting four-dimensional strongly coupled gauge theory explicitly and triggers a non-trivial renormalization Group (RG) flow from an ultraviolet (UV) fixed point to an infrared (IR) fixed point. 
The UV fixed point assured that we are in the regime where the holographic duality is best understood and the bulk metric is asymptotically AdS. The IR fixed point is needed to guarantees that the solutions are regular in the interior and the zero-temperature solution is smooth in the deep IR. This simple model has been used to study different properties such as  thermodynamics and relaxation channels \cite{Attems:2016ugt}, jet energy lost \cite{Heshmatian:2018wlv}, and entanglement entropy \cite{Rahimi:2016bbv}. 

One of the most important quantities of QGP is the energy loss of a quark moving through the plasma. An external quark was introduced in the context of AdS/CFT correspondence by adding a fundamental string attached to a flavor brane in the AdS space \cite{Herzog:2006gh,Gubser:2006bz,CasalderreySolana:2006rq}. The string endpoint specifies the heavy quark, while the string itself can be considered as a gluonic cloud around the quark. It is straightforward to show that the mass of quark is proportional to the inverse distance of the string endpoint from the boundary \cite{Herzog:2006gh}. The light quark or massless quark is mapped into a string attached to a flavor brane which is extended from the boundary to the horizon, and its dynamic in different holographic backgrounds have been studied in \cite{Chesler:2008uy,Morad:2014xla,BitaghsirFadafan:2017tci,Heshmatian:2018wlv}. On the other hand, a string attached to the boundary of AdS space is dual to an infinite mass quark. Since the first attempts to study the holographic heavy quark in $\mathcal{N} = 4$ super-Yang-Mills theory \cite{Herzog:2006gh,Gubser:2006bz}, the heavy quark dynamics have been studied in various gauge theories with gravitational duals \cite{Reiten:2019fta,Gursoy:2009kk,Horowitz:2007su,Horowitz:2015dta,Chakraborty:2014kfa,Lekaveckas:2013lha,Talavera:2006tj,Fadafan:2008uv,Giecold:2009wi,Panigrahi:2010cm,Sadeghi:2009hh,Sadeghi:2009mp,Sadeghi:2013zma,Zhang:2019cxu,Caceres:2006as,Bena:2019wcn,Atashi:2019mlo,Nakano:2006js,Matsuo:2006ws,Rougemont:2015wca,Zhang:2018rff,Zhang:2018mqt,Cheng:2014fza,NataAtmaja:2010hd}.

This paper is organised as follows: In section \ref{section:nonconBG} we briefly review the non-conformal background introduced in \cite{Attems:2016ugt}.  In section \ref{section:string} we discuss the string solution dual to a heavy quark in this non-conformal background. The numerical results of drag force and worldsheet temperature are presented in \ref{section:DF} and \ref{section:WS_T}. Section \ref{section:summary} is devoted to a summary.


\section{Non-conformal holographic model}
\label{section:nonconBG}



The action of 5-dimensional Einstein gravity coupled to a scalar field is given by
 \bea
\label{eq:action}
S=\frac{2}{\kappa_5^2} \int d^5 x \sqrt{-g} \left[ \frac{1}{4} \mathcal{R}  - \frac{1}{2} \left( \nabla \phi \right) ^2 - V(\phi) \right ] \, ,
\eea
where $\kappa_5$ is the five-dimensional Newton constant and $V(\phi)$ is a potential encoding the details of the dual gauge theory. In \cite{Attems:2016ugt}, the following potential has been considered
\bea
\label{eq:V}
L^2 V=-3 -\frac{3}{2} \phi^2 - \frac{1}{3} \phi^4 + \left( \frac{1}{3 \phi_M^2} +  \frac{1}{2 \phi_M^4}\right) \phi^6-\frac{1}{12 \phi_M^4} \phi^8 \, .
\eea
which is characterized by a single parameter, $\phi_M$. This non-trivial relatively  simple potential has a maximum at $\phi=0$ and a minimum at \mbox{$\phi=\phi_M$} corresponding to two AdS solutions at UV and IR fixed points. The radii of these two solutions are related as
\begin{equation}
\label{eq:LIR}
L_{\rm IR} = \frac{1}{1+ \frac{1}{6} \phi_M^2} L \, .
\end{equation}
Since  $L_{\rm IR} < L$, the number of degrees of freedom is smaller at IR. 

By parametrizing the vacuum metric as 
\be
\label{eq:adsvac}
ds^2 = e^{2 A(r)} \left(-d t^2 + d{\bf x}^2\right) + dr^2 \, ,
\ee
the vacuum solution can be analytically obtained for arbitrary $\phi_M$
\begin{eqnarray}
\label{eq:vacmetricsol}
e^{2 A}&=& \frac{\Lambda^2 L^2}{\phi^2} \,
  \left(1- \frac{\phi ^2}{\phi _M^2} \right)^{\frac{\phi_M^2}{6}+1} \, 
  e^{-\frac{\phi ^2}{6}}  \,,
\\[2mm]
\label{eq:phisol}
\phi(r)&=& \frac{\Lambda L \, e^{-r/L}}{\sqrt{1+ \frac{\Lambda^2 L^2}{\phi_M^2}e^{-2 r/L} }} \,.
\end{eqnarray}
It is shown that the arbitrary parameter $\Lambda$ is responsible for breaking the conformal invariance explicitly \cite{Attems:2016ugt}.

\begin{figure}
\begin{subfigure}{.5\textwidth}
  \centering
  \includegraphics[width=0.95\linewidth]{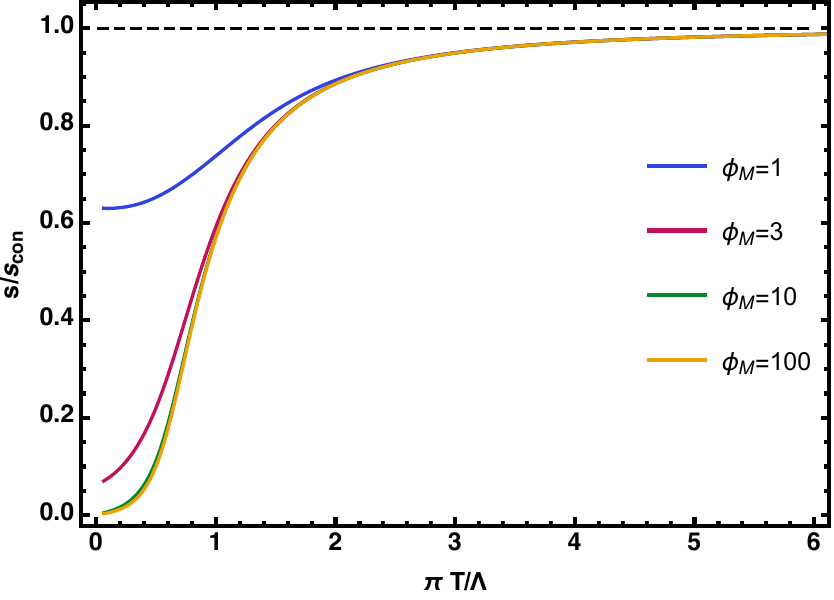}
   \caption{}
  \label{sratio}
\end{subfigure}%
\begin{subfigure}{.5\textwidth}
  \centering
  \includegraphics[width=0.97\linewidth]{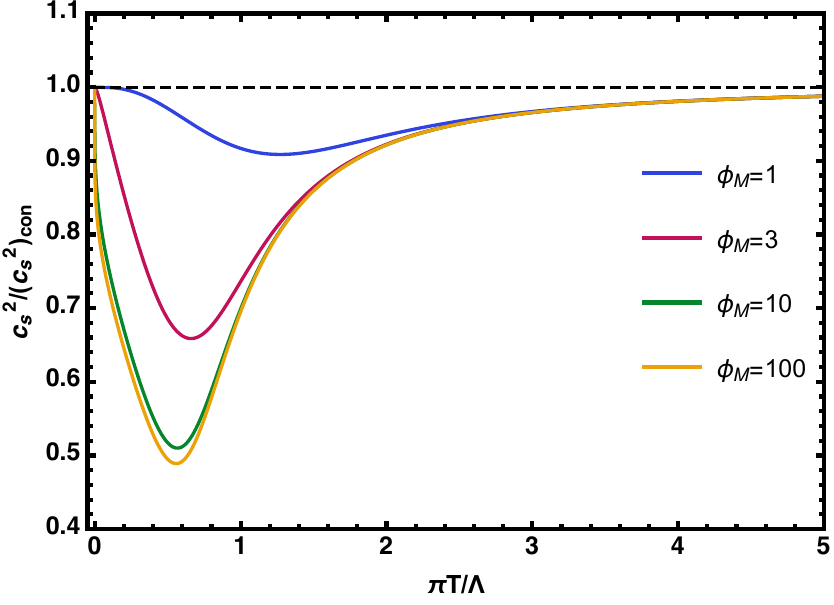}
   \caption{}
  \label{soundspeed}
\end{subfigure}%
\caption{(Color online) (\subref{sratio}) The ratio of non-conformal to the conformal entropy,(\subref{soundspeed}) the ratio of speed of sound square to its conformal limit as a function of temperature for different values of $\phiM$.}
\label{backgroundsplot}
\end{figure}

The black brane solution of action \ref{eq:action} can be calculated numerically by the following ansatz in the Eddington-Finkelstein coordinate 
\be
\label{eq:blackads}
ds^2= e^{2 A}\left(-h(\phi) d\tilde t\, ^ 2 + d{\bf x}^2 \right) -2 e^{A+B} L \, d\tilde t \,d\phi \, ,
\ee
where $\phi$ considered being the radial coordinate such that the boundary and the horizon are located at $\phi=0$ and $\phi=\phi_H$, respectively. In fact, the value of the scalar field at the horizon, $\phi_H$ characterizes the black-brane solution and its thermodynamic properties explicitly. The dual gauge theory is conformal both at the UV and at the IR. Therefore the high and low temperature behavior of thermodynamical variables must match with the relativistic conformal theory. The entropy ratio to its conformal value, $s/s_{con}$, and the ratio of speed of sound for different values of $\phi_M$ is plotted in \fig{sratio} and \fig{soundspeed}, respectively. The entropy ratio approaches one at high temperatures and $(L_{IR}/L)^3$ at low temperatures. The speed of sound reaches its conformal value both at high and low temperatures while deviating from its conformal value at intermediate temperatures, which can be interpreted as a measure of the non-conformality of the gauge theory. The detail of the numerical procedure is presented in \cite{Attems:2016ugt}. 

\section{Classical trailing string}
\label{section:string}

This section studied a heavy quark moving through an infinite volume of quark-gluon plasma with a fixed velocity $v$ at a finite temperature $T$. The heavy quark feels a drag force and consequently losses energy. In the dual theory, the dynamic of this quark can be described by an open string whose endpoint is attached to the UV boundary of AdS space and moves at the constant speed of $v$ with a string tail into the AdS bulk \cite{Herzog:2006gh,Gubser:2006bz}.

The dynamics of the string is described by the Nambu-Goto action
\begin{equation}
S_{NG} =\,{1 \over 2 \pi \alpha'} \int d\sigma d \tau \, \sqrt{ -\det \gamma_{\alpha\beta}},
\label{eq:NG}
\end{equation}
where $\alpha'=l_s^2$ is the square of string's fundamental length and $\gamma_{\alpha\beta}$ is the induced metric of the world-sheet defined as 
\begin{equation}
\gamma_{\alpha\beta} =\,G_{MN}\, \partial_{\alpha} X^{M}\, \partial_{\beta} X^{M}\, ,
\label{eq:Induced_metric}
\end{equation}
where $G_{MN}$ are the metric component of \eqq{eq:blackads} in the Poincare coordinate
\begin{equation}
G_{tt}(\phi)=-\,h(\phi)\, e^{2A(\phi)} \,\,\,, G_{xx}(\phi)= e^{2A(\phi)}\,\,\,,G_{\phi \phi}(\phi)=\frac{L^2}{h(\phi)} e^{2B(\phi)}\,.
\label{eq:metric}
\end{equation}
Henceforward we assume $L=1$ for simplicity. By considering the following ansatz for the string embedding function in the static gauge $(\tau=t\,,\sigma=\phi)$ 
\begin{equation}
X^M(\tau,\sigma) = (t,x=v t+\xi(\phi),0,0,\phi) \, ,
\label{eq:embeddingf}
\end{equation}
the induced metric on the world-sheet becomes
\begin{equation}
\gamma_{\alpha\beta}= e^{2A}  \left(\begin{array}{cc} 
-(h-v^2)   & v\, \xi'\\
v \,\xi'  & \frac{e^{2(B-A)}}{h}+\xi'^2
\\ \end{array}\right) \, ,
\label{eq:inducedmetric}
\end{equation}
leads to the corresponding Lagrangian
\begin{equation}
\mathcal{L}=-\frac{e^{A(\phi )}}{2 \pi  \alpha' } \sqrt{ e^{2 A(\phi )} h(\phi ) \xi '(\phi )^2-\frac{v^2 e^{2 B(\phi )}}{h(\phi )}+e^{2 B(\phi )}}\,.
\label{eq:Lag}
\end{equation}
Since the Lagrangian does not depend on the $\xi$ explicitly, the string equation of motion simplifies as
\begin{equation}
\frac{1}{2 \pi  \alpha' }\frac{e^{2 A(\phi )} h(\phi ) \xi '(\phi )}{\sqrt{e^{2 (B(\phi )-A(\phi ))}-v^2\,\frac{ e^{2 (B(\phi )-A(\phi ))}}{h(\phi )}+h(\phi ) \xi '(\phi )^2}}\,=\,const \equiv \pi_{\xi} \,,
\label{eq:conjMom}
\end{equation}
where $\pi_\xi$ is the constant of motion. From above, the equation for $\xi'$ can be obtained as
\begin{equation}
\xi'(\phi)=\pm \frac{2\pi \alpha' \pi _{\xi } e^{B(\phi )-A(\phi )}}{h(\phi )}\, \sqrt{\frac{h(\phi )-v^2}{e^{4 A(\phi )} h(\phi )-\left(2\pi \alpha' \pi _{\xi }\right)^2}}\, .
\label{eq:xip}
\end{equation}
The numerator and denominator inside the square root are positive near the boundary and negative around the horizon. Requiring the real values for the string profile, $\xi'$ means that the numerator and the denominator have to change their sign at the same point denoted by $\phi_s$
\begin{eqnarray}\label{phis}
h(\phi_s )-v^2=0\, ,\nonumber \\
e^{4 A(\phi_s )} h(\phi_s )-\left(2\pi \alpha' \pi _{\xi }\right)^2=0\, .
\end{eqnarray}
From the above equations, the constant of the equation of motion can be determined as
\begin{eqnarray}\label{pi}
 \pi _\xi= \frac{1}{2\pi \alpha'} \,v \,e^{2 A(\phi_s)}  \,.
\end{eqnarray}
Substituting the $\pi_\xi$ into Eq. (\ref{eq:xip}) yields to 
\begin{equation}
\xi'(\phi)=\pm \frac{ v e^{2A(\phi_s )}  e^{B(\phi )-A(\phi )}}{h(\phi )}\, \sqrt{\frac{ h(\phi)-h(\phi_s)}{e^{4 A(\phi )} h(\phi )-e^{4 A(\phi_s )} h(\phi_s)}} \, .
\label{eq:xip2}
\end{equation}
The square root is now well-defined for the region outside the horizon $(0<\phi<\phi_H)$, and this equation can be solved numerically to obtain the string profile. 

\subsection{Drag force}
\label{section:DF}

The drag force, which indicates the dissipation of quark momentum into the plasma, is given by
\begin{equation}
\label{DragF}
F_{drag}= -\pi^1_x  \, ,
\end{equation}
where $\pi^1_x $ is the momentum that is lost by flowing from the string to the horizon. 
The canonical momentum densities associated with the string can be obtained from varying the action with respect to the derivatives of the embedding functions
\begin{eqnarray}
    \left( \begin{array}{c}
  \label{densities}
    \pi^0_t \\ \pi^0_x 
    \end{array}
    \right)
   & =&  \frac{ 1 }{2\pi \alpha'} 
    \frac{e^{4A(\phi)} }{\sqrt{ -\det \gamma_{\alpha\beta}}} \,
    \left(
    \begin{array}{c}
   - e^{2\left(B(\phi)-A(\phi)\right)}-h(\phi)\, \xi'(\phi)^2 \\
   e^{2\left(B(\phi)-A(\phi)\right)}\, v\,/\,h(\phi)  \end{array}
    \right) ,
     \\ \nonumber
     \\ 
    \left(
  \begin{array}{c}
   \label{flows}
    \pi^1_t \\ \pi^1_x 
    \end{array}
    \right)
  &  =&
    \frac{ 1 }{2\pi \alpha'} 
    \frac{e^{4A(\phi)} }{\sqrt{ -\det \gamma_{\alpha\beta}}} \,
        \left(
    \begin{array}{c}
   v \,\xi '(\phi) \,h(\phi)\\
   - \xi '(\phi) \,h(\phi)
    \end{array}
    \right).
\end{eqnarray}
Integrating the $\pi^0_{t}$ and $\pi^0_{x}$ along the string gives us the total energy and the total momentum in the direction of motion of the string, respectively. While, $\pi^1_{t}$ and $\pi^1_{x}$ are the energy and momentum flow along the string and similar to the case of $\mathcal{N}=4$ SYM plasma, $ \pi^1_t =-v\, \pi^1_x $. 
By substituting the string solution, \eqq{eq:xip2} in the \eqq{flows}, it is evident that the $\pi^1_{x}$ equals the constant of the equation of motion, $\pi_{\xi}$. This means that if we pull the quark with a constant velocity, the fraction of energy flows at a given point along the string, $ \pi^1_t$ is constant. Using the \eqq{pi}, and \eqq{DragF} the drag force is 
\begin{equation}\label{DragFs}
F_{drag}=-\frac{1} {2\pi \alpha'}\,v\,e^{2 A(\phi_s)} \, ,
\end{equation}
which is the momentum flow along the string at point $\phi_s$. If we use the metric of AdS-Sch, the drag force for $\mathcal{N}=4$ SYM plasma can be obtained as \cite{Herzog:2006gh,Gubser:2006bz,CasalderreySolana:2006rq} 
\begin{equation}\label{DragF_C}
F_{drag}^{SYM}=-\frac{\pi\,T^2\,\sqrt{\lambda}}{2} \frac{v}{\sqrt{1-v^2}}   \, ,
\end{equation}
where we have used the relation $L^4=\lambda\,\alpha'^2$. In the case of non-conformal background, the first equation of \eqq{phis} should be solved numerically to obtain the $\phi_s$ and then the drag force can be calculated using \eqq{DragFs}. 

\begin{figure}
\begin{subfigure}{.5\textwidth}
  \centering
  \includegraphics[width=0.99\linewidth]{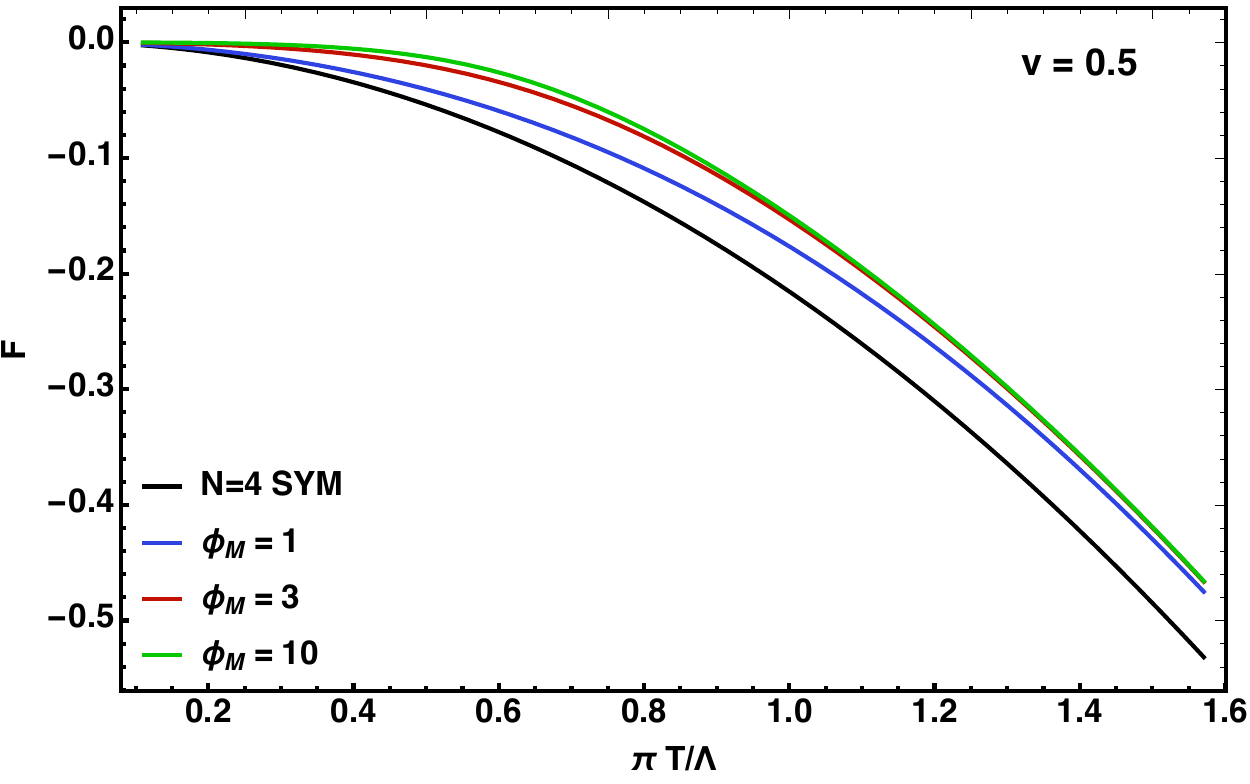}
   \caption{}
  \label{F(0.5)}
\end{subfigure}%
\begin{subfigure}{.5\textwidth}
  \centering
  \includegraphics[width=0.99\linewidth]{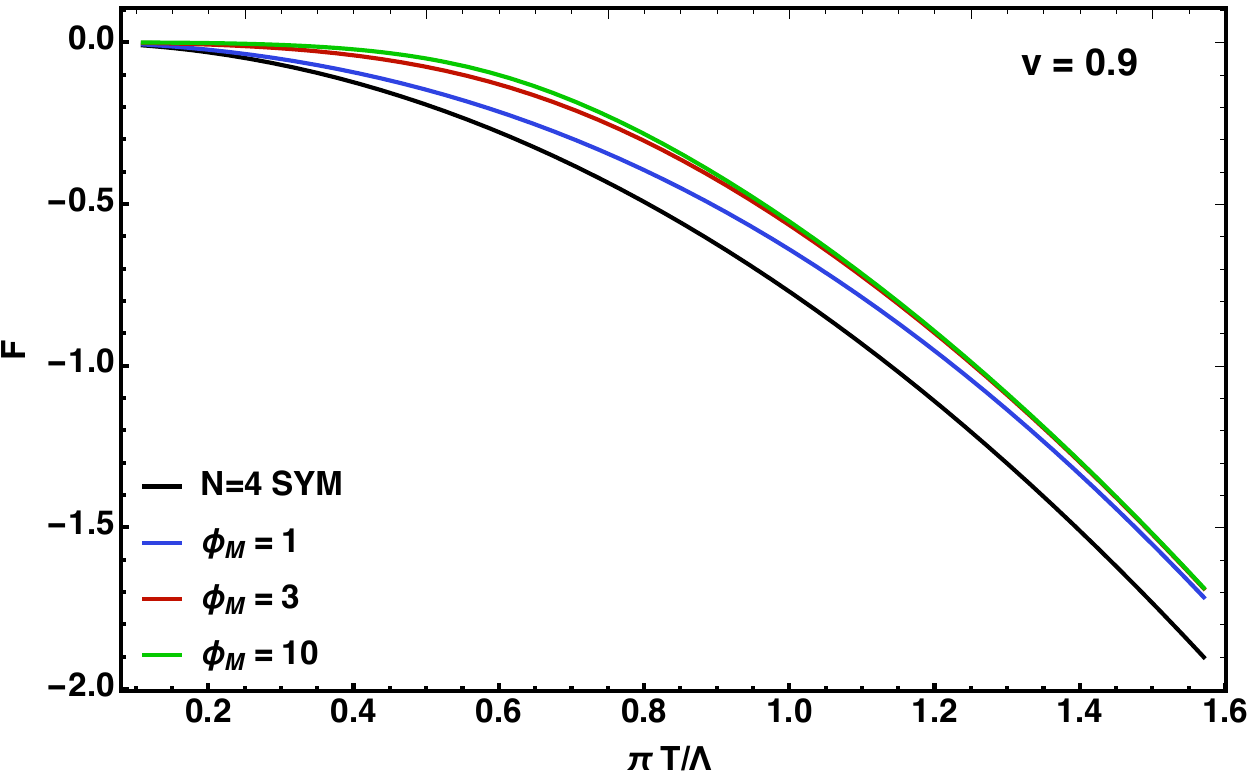}
   \caption{}
  \label{F(0.9)}
\end{subfigure}%
\caption{(Color online) Drag force as a function of temperature for different values of $\phi_M$ compared with the drag force in $\mathcal{N}=4$ SYM.}
\label{F-T}
\end{figure}

\begin{figure}
\begin{subfigure}{.5\textwidth}
  \centering
  \includegraphics[width=0.99\linewidth]{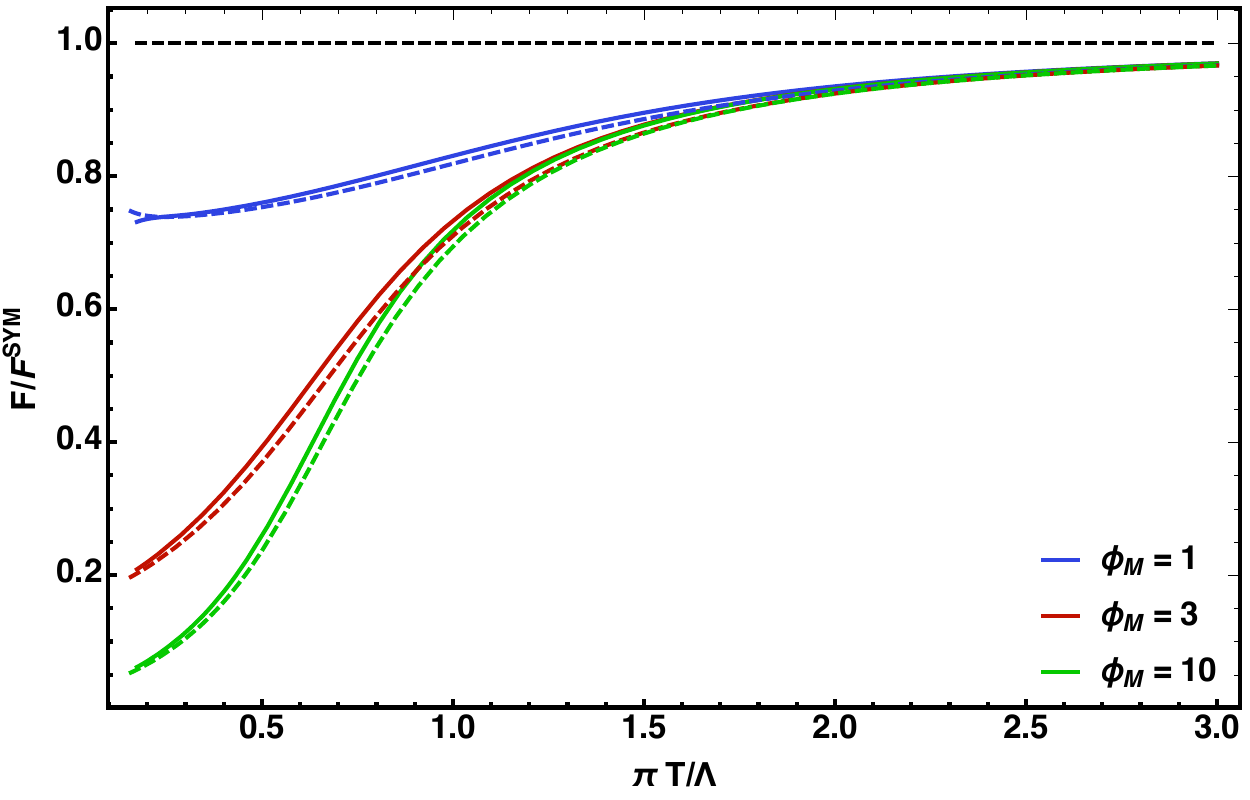}
   \caption{}
  \label{Fratio1}
\end{subfigure}%
\begin{subfigure}{.5\textwidth}
  \centering
  \includegraphics[width=0.99\linewidth]{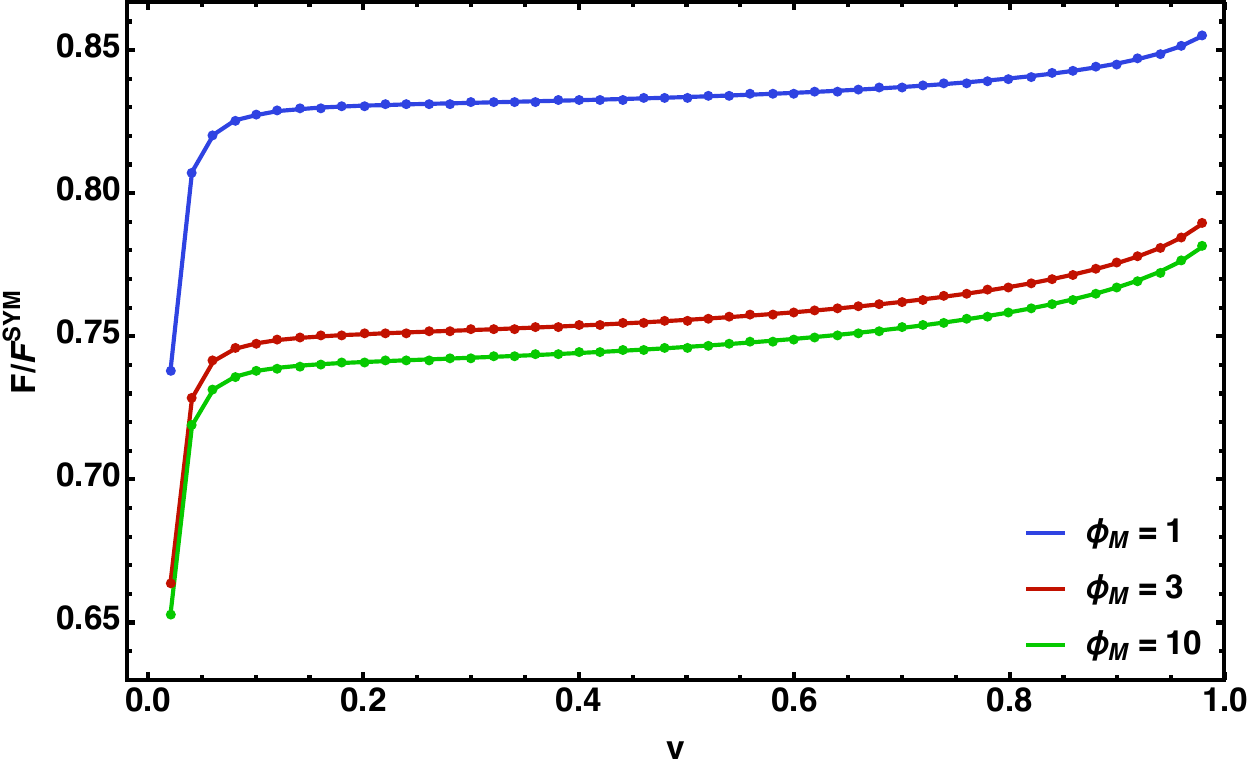}
   \caption{}
  \label{F-v}
\end{subfigure}%
\caption{(Color Online) The ratio of drag force to its conformal value for different values of $\phi_M$ as a function of  (\subref{Fratio1}) temperature, (\subref{F-v}) velocity. The solid and dashed lines in (a) represent $v=0.9$ and $v=0.5$, respectively. The temperature of both plasma in (b) sets to 350 MeV.}
\label{Fratio}
\end{figure}

Our numerical results for the drag force as a function of temperature for two values of quark velocity are plotted in Fig. \ref{F-T}. The drag force in this non-conformal plasma is smaller than the drag force in $\mathcal{N}=4$ SYM plasma with the same temperature and decreases by increasing the degree of non-conformality, $\phi_M$. Also, the drag force increases by increasing the quark velocity. In Fig. \ref{Fratio1}, the drag force ratio to its conformal limit is plotted for different values of $\phi_M$ as a function of temperature. The solid and dashed lines represent the drag force for $v=0.9$, and $v=0.5$ respectively. As is evident from this figure, at high temperatures, where the background reaches its conformal limit, the ratio of drag force tends to one. 

In addition, the ratio of drag force to its conformal value vs velocity is presented for different values of $\phi_M$ in Fig. \ref{F-v} to explore the velocity dependency. The temperature in both $\mathcal{N}=4$ SYM plasma and non-conformal plasma was set to 350 MeV. We find that for all values of $\phi_M$ the ratio is less than one at the fixed temperature but increases slowly by increasing the velocity of quark. 

\subsection{World-sheet temperature}
\label{section:WS_T}

One can reparametrize the world-sheet coordinate, \eqq{eq:embeddingf} as follows
 \begin{eqnarray}\label{new_coordinate}
\tau&=&t + K(\phi)\, , \nonumber \\
x&=&v t +v K(\phi)+\xi(\phi) \, .
\end{eqnarray}
Therefore, the background induced metric, \eqq{eq:inducedmetric} becomes
\begin{equation}
\gamma_{\alpha\beta}= e^{2A}  \left(\begin{array}{cc}
-(h-v^2)   & v\, (\xi'+v\,K')-h\,K'\\
v\, (\xi'+v\,K')-h\,K' &\,\,\, \frac{e^{2(B-A)}}{h}+(\xi'+v\,K')^2-h\,K'^2
\\ \end{array}\right)\, .
\label{eq:inducedmetric2}
\end{equation}
By choosing a particular ansatz of the form of
 \begin{eqnarray}\label{Kp}
K'(\phi)=\frac{v\, \xi'}{h-v^2}\, ,
\end{eqnarray}
and using \eqq{pi}, and \eqq{eq:xip2}, the induced metric diagonalizes as 
\begin{equation}
\gamma_{\alpha\beta}= e^{2A(\phi)}  \left(\begin{array}{cc}
-(h(\phi)-v^2)   & 0  \\
0& \frac{e^{2(B(\phi)+A(\phi))}}{e^{4A(\phi)}\,h(\phi)- e^{4A(\phi_s)}\,v^2}
\\ \end{array}\right) \, ,
\label{eq:inducedmetric3}
\end{equation}
which describe the metric of a two-dimensional world-sheet blackhole with a horizon radius of $\phi_s$. The Hawking temperature of the world-sheet blackhole indicated as $T_{wsh}$ is
 \begin{eqnarray}\label{Twsh}
T_{wsh}=\frac{1}{4\pi}\sqrt{e^{2A(\phi_s)-2B(\phi_s)}\,h'(\phi_s)( 4\,v^2\,A'(\phi_s)+h'(\phi_s)) }\, .
\end{eqnarray}
In the conformal limit, AdS-Schwarzschild metric, the world-sheet temperature depends on the blackhole temperature and the velocity of quark as
\begin{eqnarray}\label{Twsh_con}
T_{wsh}^{conf}=\frac{T}{\sqrt{\gamma}}\, .
\end{eqnarray}

In Fig. \ref{Twsh}, the ratio of world-sheet temperature to its conformal value is plotted in terms of the plasma temperature for different values of $\phi_M$. The world-sheet temperature decreases by increasing the value of $\phi_M$. The deviation from conformal behaviour is dominant at intermediate temperature. At high temperatures, the ratio tends to one, as expected. 
\begin{figure}
  \centering
  \includegraphics[width=0.95\linewidth]{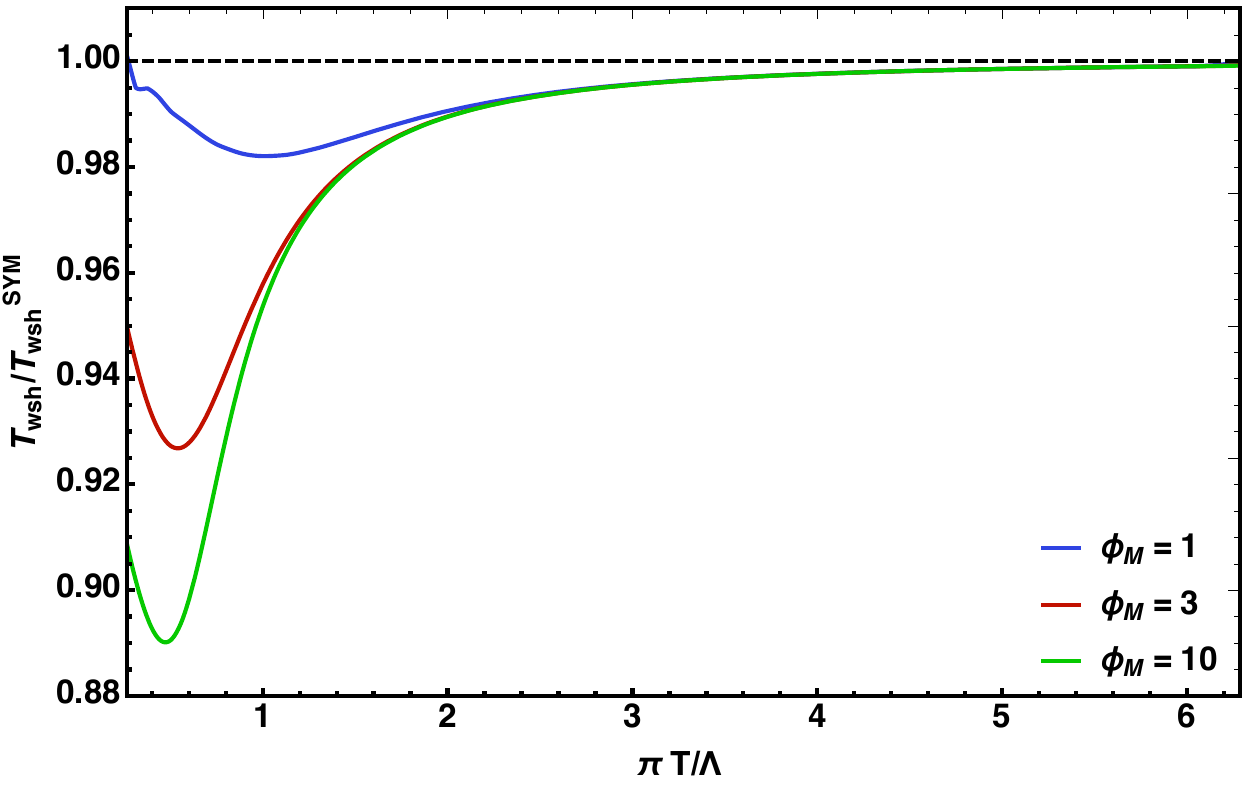}
   \caption{(Color Online) The ratio of world-sheet temperature to its conformal value in terms of the plasma temperature for different values of $\phi_M$.}
  \label{Twsh}
\end{figure}


\section{Summary}
\label{section:summary}

This paper studied the dynamics of a heavy quark moving through a strongly coupled plasma with broken conformal symmetry using the gauge/gravity duality. 
 We considered a holographic five-dimensional model consisting of Einstein gravity coupled to a scalar field with a non-trivial potential corresponding to a dual four-dimensional non-conformal gauge theory that exhibits a renormalization group flow between two different types of fixed points (located at UV and IR) at zero temperature \cite{Attems:2016ugt}. The parameter $\phi_M$ indicates the deviation from conformality, as shown in Fig. \ref{backgroundsplot}. According to the AdS/CFT dictionary, the heavy quark is associated with a string attached to the boundary of the AdS space. The string equation of motion was solved numerically for different values of parameter $\phi_M$, and then the drag force was computed by obtaining the corresponding conjugate momenta. Our results indicate that the drag force in this non-conformal plasma is smaller than $\mathcal{N} = 4$ super-Yang-Mills plasma. By increasing the level of non-conformality (the value of parameter $\phi_M$), the drag force decreases, see Fig. \ref{F-T}. The ratio of drag force to its conformal limit reaches one at high temperatures. At finite temperature, this ratio is smaller than one for different values of $\phi_M$ but increases slowly by increasing the velocity of quark, see Fig. \ref{Fratio}.

The world-sheet horizon was also calculated, and it was shown that the drag force is equivalent to the energy flow at the world-sheet horizon. The corresponding world-sheet temperature has been calculated and compared with its conformal limit. Results confirm that the deviation from the conformal limit is dominated at the intermediate plasma temperature.


\acknowledgments
The authors would like to thank Dr. Can Kozçaz for his useful discussion and comments. R. Morad and T. Domurcukgul acknowledge the Scientific and Technological Research Council of Turkey (Tübitak), the Department of Science Fellowships and Grant Programs (BIDEB) for the fellowship program for visiting scientists (2221) and ARDEB 1001 – Support Program for Scientific and Technological Research Projects, respectively.

\end{document}